\documentclass[reprint,aps]{revtex4-1}
\usepackage[utf8]{inputenc}
\usepackage[T1]{fontenc}
\usepackage{physics}
\usepackage{amsmath}
\usepackage{amssymb}
\usepackage{amsthm}
\usepackage{tikz}
\theoremstyle{definition}
\newtheorem*{exmp*}{Example}
\pgfdeclarelayer{background}
\pgfdeclarelayer{foreground}
\pgfsetlayers{background,main,foreground}
\begin{document}
\title{Quantum computation from fermionic anyons on a 1D lattice}
\author{Allan D. C. Tosta}
\author{Daniel J. Brod}
\author{Ernesto F. Galv\~{a}o}
\affiliation{Instituto de F\'isica, Universidade Federal Fluminense, Av. Gal. Milton Tavares de Souza s/n, Niter\'oi, RJ, 24210-340, Brazil}
\date{\today}

\begin{abstract}
Fermionic linear optics corresponds to the dynamics of free fermions, and is known to be efficiently simulable classically. We define fermionic anyon models by deforming the fermionic algebra of creation and annihilation operators, and consider the dynamics of number-preserving, quadratic Hamiltonians on these operators. We show that any such deformation results in an anyonic linear optical model which allows for universal quantum computation.
\end{abstract}
\maketitle

\section{Introduction}

There are good reasons to study computational models using particles with some kind of fractional statistics. Experimentally, there has been a growing interest in finding such exotic particles in condensed matter systems such as the one exhibiting the fractional quantum hall effect (FQHE) \cite{Laughlin1983}, as they can simplify the description of a multitude of complex condensed matter phenomena. They can also be used as a platform for models of quantum computation that are intrinsically robust to decoherence under specific conditions \cite{Nayak1996b,Kitaev2003,Kitaev2006a,Nayak2008a}. Theoretically, understanding the computational power of models based on such particles can give us more knowledge about the transition from classical to quantum computation and what resources make this transition possible. 

In this paper we generalize the model of quantum computation based on the dynamics of non-interacting fermions (known as fermionic linear optics (FLO) \cite{Valiant2001a,Terhal2002,Bravyi2004}) to one of anyons on a 1-D lattice \cite{Amico1998One-dimensionalStatistics,Batista2008}. This anyon system is defined via a deformation of the fermionic anti-commutation relations that introduces non-trivial exchange phases (also called fractional exchange phases).

A system of identical particles with abelian fractional exchange statistics is one where the multiparticle wave function gains an arbitrary complex phase factor under an operation that exchanges particle positions \cite{Mirman1973,Leinaas1977}. A non-abelian system is very similar but, in this case, each particle has an internal Hilbert space, and because of it the exchange phases are replaced by arbitrary unitary matrices. Particles with these properties are called abelian anyons in the first case and non-abelian anyons in the second \cite{Wilczek1982}. Such statistics have a topological origin related to the dimensionality of the physical space, and can be observed in two-dimensional many-body systems subject to a process know as transmutation of statistics \cite{Wilczek1982}. This involves many-body interactions described by non-dynamical effective vector potentials, whose only roles are to associate fictitious magnetic fluxes to charged particles, generating the exchange phase factor via the Aharonov-Bohm effect. These fictitious fields are described by topological quantum field theories (TQFT), such as the Chern-Simons theory \cite{Witten1988}, and used to describe many classes of states of FQHE systems \cite{Frohlich1991,Frohlich1991a}.

In contrast to this topological description, known to be valid only in two dimensions, a system is said to have fractional exclusion statistics if the maximum number of particles allowed per mode is a finite integer not equal to one \cite{Haldane1991}. This is a dimension independent generalization of the Pauli Exclusion Principle which also applies to the description of some subsets of FQH states that possess exchange statistics. This generalization is unrelated to statistical transmutation and is, in fact, independent of it. For such a definition, spinless fermions and spinless hard-core bosons are both seen as fermions, since both obey the exclusion principle, even though their commutation relations differ if particles are in different states. 

The type of anyonic system we will be concerned with here is defined by modified commutation relations, referred to in the literature as deformed commutation relations \cite{Greenberg1990,Batista2008,Meljanac1994,Bardek1994a,Amico1998One-dimensionalStatistics}, obtainable from the conventional fermionic and bosonic ones via specific mappings \cite{Meljanac1994b,ElBaz2003,Girardeau2006}. These mappings, which have a similar form to the Jordan-Wigner transformation \cite{Jordan1928}, relate fermionic and bosonic operators on a lattice to corresponding anyonic operators on the same lattice.

Systems defined in terms of deformed commutation relations allow for the study of anyonic gases and other analogues of one-dimensional condensed matter systems using conventional many-body quantum mechanics and renormalization group methods \cite{Amico1998One-dimensionalStatistics,Hao2008,Hao2012,Tang2015}. Besides new effects in thermodynamics and the study of phase transitions \cite{Keilmann2011,Liu2018AsymmetricStatistics,Wang2014}, these systems are interesting also due to practical proposals for implementations e.g.\ in optical lattices  \cite{Greschner2015,Cardarelli2016,Strater2016FloquetLattice}, and even a proposal for simulation using classical optics \cite{Longhi2012AnyonsRealization}.

We define a computational model similar to the Boson Sampling model \cite{Aaronson2010} where a finite number of photons are input into a circuit made of optical devices such as beam-splitters and phase-shifters, followed by photon-counting detectors. The anyonic model used in this work interpolates between one-dimensional fermions and one-dimensional hard-core bosons. We  remark that our proposal is different from the one made in topological quantum computing (TQC), which deals with the computational power of braiding non-abelian anyons and is intrinsically fault-tolerant \cite{Freedman2000,Kitaev2003,Bravyi2005a,Cui2015,Cui2015a,Levaillant2015} .

This paper is organized as follows. In Section \ref{review} we define and review the properties of several computational models based on quadratic number-preserving particle dynamics, including FLO, hard-core bosons, and the anyonic model we investigate \cite{Wu2002,Kempe2002}. In Section \ref{anyon1} we solve the general two-particle dynamics and define equivalent ``linear-optical'' devices, highlighting the differences with respect to its bosonic and fermionic counterparts. In Section \ref{power} construct a universal set of gates inspired by the one obtained in \cite{Kempe2002}, showing that our model is universal for quantum computation (as could be expected from the literature on extensions of FLO and related models \cite{Brod2011,Oszmaniec2017UniversalOperations}). Finally, in Section \ref{conclusion} we offer some concluding remarks.

\section{Review}\label{review}

In this Section we review how to describe non-interacting bosons and fermions, whose dynamics are respectively known as bosonic (section \ref{boson}) and fermionic linear optics (section \ref{fermion}). In section \ref{hboson} we review how qubits can be understood as a system of bosons under a hard-core interaction. In section \ref{anyon} we define the anyon model we use and how it can be constructed from deformed fermionic commutation relations, though we defer the discussion of their dynamics to a subsequent section.

%In this section we define what we mean by bosonic and fermionic linear optics, highlighting their differences, in particular with respect to computational power.  We also review how qubits can be understood as a system of bosons with a hard-core interaction. We will then review what we linear optics for this system, showing their relation with the computational power of spin Hamiltonians and in particular with the XY-interaction.

\subsection{Linear Optics}\label{boson}
Photons are described by their degrees of freedom (which we call \textit{modes}, for short).
We consider a system with $m$ modes described by creation and annihilation operators ${b}^{\dagger}_{i}$ and ${b}_{i}$ with $i=1,...,m$ which respectively, create and destroy a single photon in mode $i$. These operators obey the canonical bosonic commutation relations
\begin{subequations}\label{eq1}
\begin{align}
&{b}_{i}{b}^{\dagger}_{j}-{b}^{\dagger}_{j}{b}_{i}=\delta_{ij},\label{eq1a}\\
&{b}_{i}{b}_{j}-{b}_{j}{b}_{i}=0,\label{eq1b}\\
&{b}^{\dagger}_{i}{b}^{\dagger}_{j}-{b}^{\dagger}_{j}{b}^{\dagger}_{i}=0\label{eq1c}.
\end{align}
\end{subequations}
for all modes $i,j$. The basis vectors for this system's Hilbert space can be chosen to be
\begin{equation}
\ket{{n}^{B}_{1},...,{n}^{B}_{m}}=\frac{({b}^{\dagger}_{1})^{{n}^{B}_{1}}...({b}^{\dagger}_{m})^{{n}^{B}_{m}}}{\sqrt{{n}^{B}_{1}!...{n}^{B}_{m}!}}\ket{0_{B}},   
\end{equation}
where ${n}^{B}_{i}$ is the eigenvalue of the number operator ${N}^{B}_{i}={b}^{\dagger}_{i}{b}_{i}$. This basis is called the \textit{Fock basis} or the \textit{occupation number basis}, and we use this basis throughout this work. 

As we will shortly see, any free-particle bosonic dynamics can be expressed in terms only of one- and two-mode passive linear optical elements (as described in e.g.\  \cite{Prasad1987}). A \textit{phase-shifter} is a single-mode passive linear device described as the time-evolution operator ${PS}_{i}(\theta)=\exp{i\theta{H}^{PS}_{i}}$ of the Hamiltonian ${H}^{PS}_{i}={N}^{B}_{i}$. Its effect on Fock states results in an occupation-number dependent phase:
\begin{equation}
{PS}_{i}(\theta)\ket{{n}^{B}_{1},...,{n}^{B}_{m}}=e^{i\theta {n}^{B}_{i}}\ket{{n}^{B}_{1},...,{n}^{B}_{m}},
\end{equation}
a phase-shifter acts on creation operators as:
\begin{equation}
{PS}_{i}(\theta){b}^{\dagger}_{j}{PS}_{i}(-\theta)=e^{i\theta\delta_{ij}}{b}^{\dagger}_{j}.
\end{equation}
    
A \textit{beam-splitter} is a two-mode passive linear device described by the time-evolution operator ${BS}_{ij}(\theta)=\exp{i\theta{H}^{BS}_{ij}}$, corresponding to the Hamiltonian ${H}^{BS}_{ij}={b}^{\dagger}_{i}{b}_{j}+{b}^{\dagger}_{j}{b}_{i}$. Its effect in creation operators is given by the matrix equation
\begin{equation}\label{eq5}
{BS}_{ij}(\theta)
\begin{bmatrix}
{b}^{\dagger}_{i}\\
{b}^{\dagger}_{j}
\end{bmatrix}
{BS}_{ij}(-\theta)=
\begin{bmatrix}
\cos{\theta} & i\sin{\theta}\\
i\sin{\theta} & \cos{\theta}
\end{bmatrix}
\begin{bmatrix}
{b}^{\dagger}_{i}\\
{b}^{\dagger}_{j}
\end{bmatrix}
.
\end{equation}

Another way to understand how these devices act is to consider a system with a single photon which can be in any of the $m$ modes. The basis states of this system in the occupation number representation are the $m$ states $\ket{1,0,0,..,0},\ket{0,1,0,...,0},...,\ket{0,0,0,...,1}$. In terms of these states, the actions of beam-splitters and phase-shifters are $SU(2)$ matrices in the subspaces on which they act. In fact it was proven that any $SU(m)$ matrix can be constructed in this single photon system using only successive applications of phase-shifters and beam-splitters \cite{Reck1994}. The computational model where the computational basis states are the Fock states, the unitaries are arbitrary circuits of phase-shifters and beam-splitters between two arbitrary modes, and measurements are made in the Fock basis is called the Boson Sampling model \cite{Aaronson2010}. This model  is not known to be universal for quantum computational, but there is evidence that it is hard to simulate on a classical computer given some complexity-theoretic assumptions.

\subsection{Fermionic Linear Optics}\label{fermion}
We now turn to the \textit{fermionic linear optics} (FLO) model as described e.g in \cite{Terhal2002,Bravyi2002,Valiant2001a}, and specialize it to our needs. Consider a system of abstract fermionic modes described by creation and annihilation operators ${f}^{\dagger}_{i}$ and ${f}_{i}$ on $m$ modes, satisfying the \textit{canonical fermionic anti-commutation relations}
\begin{subequations}\label{eq6}
\begin{align}
&{f}_{i}{f}^{\dagger}_{j}+{f}^{\dagger}_{j}{f}_{i}=\delta_{ij},\label{eq6a}\\    
&{f}_{i}{f}_{j}+{f}_{j}{f}_{i}=0,\label{eq6b}\\
&{f}^{\dagger}_{i}{f}^{\dagger}_{j}+{f}^{\dagger}_{j}{f}^{\dagger}_{i}=0,\label{eq6c}
\end{align}
\end{subequations}
for all modes $i,j$. The occupation number operator is ${N}^{F}_{i}={f}^{\dagger}_{i}{f}_{i}$ and the vacuum state is $\ket{0_{F}}$. The Fock basis for fermions comprises basis states
\begin{equation}
\ket{{n}^{F}_{1},...,{n}^{F}_{m}}=({f}^{\dagger}_{1})^{{n}^{F}_{1}}...({f}^{\dagger}_{m})^{{n}^{F}_{m}}\ket{0_{F}},
\end{equation}
where the ${n}^{F}_{j}$ are eigenvalues of the corresponding number operators which, due to the commutation relations, can only be $0$ or $1$. 

We call \textit{passive fermionic linear optical elements}, unitaries of the form
\begin{subequations}
\begin{align}
{PS}_{i}(\theta)&=\exp{i\theta {H}^{PS}_{i}},\\
{BS}_{ij}(\theta)&=\exp{i\theta {H}^{BS}_{ij}},
\end{align}
with hamiltonians given by
\begin{align}
{H}^{PS}_{i}&={N}^{F}_{i},\\
{H}^{BS}_{ij}&={f}^{\dagger}_{i}{f}_{j}+{f}^{\dagger}_{j}{f}_{i}.
\end{align}
\end{subequations}
The action of these elements over creation operators is very similar to the bosonic case, and is given by the equations
\begin{subequations}
\begin{align}
{PS}_{i}(\theta){f}^{\dagger}_{j}{PS}_{i}(-\theta)&=e^{i\theta\delta_{ij}}{f}^{\dagger}_{j},\\
{BS}_{ij}(\theta)\label{eq9b}
\begin{bmatrix}
{f}^{\dagger}_{i}\\
{f}^{\dagger}_{j}
\end{bmatrix}
{BS}_{ij}(-\theta)&=
\begin{bmatrix}
\cos{\theta} & i\sin{\theta}\\
i\sin{\theta} & \cos{\theta}
\end{bmatrix}
\begin{bmatrix}
{f}^{\dagger}_{i}\\
{f}^{\dagger}_{j}
\end{bmatrix}
.
\end{align}
\end{subequations}
As in the bosonic case, one can construct any $SU(m)$ operator over the Hilbert space of a single fermion in $m$ modes using only compositions of beam-splitters and phase-shifters.

By taking as computational basis states the occupation number basis, and as logic gates fermionic linear optical elements between arbitrary modes, and measurements in the occupation number basis, we define the computational model called Fermionic Linear Optics. This model of computation is proven to be exactly efficiently simulable by a classical computer in \cite{Terhal2002}, which means that there is an efficient classical algorithm to evaluate $\abs{\bra{y}U\ket{x}}^{2}$ for arbitrary computational states $\ket{x}$ and $\ket{y}$ and arbitrary unitaries generated by the application of passive optical elements (and any desired marginal probabilities).

\subsection{Qubits as hard-core bosons}\label{hboson}
To make a point of comparison we now give an alternate description of the usual circuit model of computation on qubits. We  will call this model \textit{qubit linear optics}. Consider a system of $m$ qubits, with the Pauli matrices denoted by $X_{i},Y_{i},Z_{i}$. Following \cite{Wu2002} we can write the usual computational basis states as a Fock space basis given by the equations
\begin{equation}
\ket{{n}^{Q}_{1},...,{n}^{Q}_{m}}=({\pi}^{\dagger}_{1})^{{n}^{Q}_{1}}...({\pi}^{\dagger}_{m})^{{n}^{Q}_{m}}\ket{0_{Q}},
\end{equation}
with the creation and annihilation operators given by 
${\pi}^{\dagger}_{i}=\frac{1}{2}(X_{i}+iY_{i})$ and
${\pi}_{i}=\frac{1}{2}(X_{i}-iY_{i})$ acting on state $\ket{0_{Q}}=\ket{0,...,0}$, and number operator ${N}^{Q}_{i}={\pi}^{\dagger}_{i}{\pi}_{i}=\frac{1}{2}(\mathbf{1}_{i}+Z_{i})$ whose eigenvalues ${n}^{Q}_{i}$ can be only $0$ or $1$. These operators must obey the algebra
\begin{subequations}
\begin{align}
&{\pi}_{i}{\pi}^{\dagger}_{j}-{\pi}^{\dagger}_{j}{\pi}_{i}=0,\\
&{\pi}_{i}{\pi}_{j}-{\pi}_{j}{\pi}_{i}=0,\\
&{\pi}^{\dagger}_{i}{\pi}^{\dagger}_{j}-{\pi}^{\dagger}_{j}{\pi}^{\dagger}_{i}=0,
\end{align}
for all modes $i,j$ with $i\neq j$, [compare with Eqs.\ (\ref{eq1})] and
\begin{align}
&{\pi}_{i}{\pi}^{\dagger}_{i}+{\pi}^{\dagger}_{i}{\pi}_{i}=1\label{eq11d},\\
&({\pi}^{\dagger}_{i})^{2}=({\pi}_{i})^{2}=0,\label{eq11e}
\end{align}
for each mode $i$ [compare with Eqs.(\ref{eq6})].
\end{subequations}
This algebra forbids more than one boson in the same mode, which is why this system can be understood as bosons with a hard-core interaction.

Similarly to bosonic and fermionic linear optics, qubit passive linear optical elements are defined as unitary operators of the form
\begin{subequations}
\begin{align}
{PS}_{i}(\theta)&=\exp{i\theta {H}^{PS}_{i}},\\
{BS}_{ij}(\theta)&=\exp{i\theta {H}^{BS}_{ij}},
\end{align}
with hamiltonians given by
\begin{align}
{H}^{PS}_{i}&={N}^{Q}_{i},\\
{H}^{BS}_{ij}&={\pi}^{\dagger}_{i}{\pi}_{j}+{\pi}^{\dagger}_{j}{\pi}_{i}.
\end{align}
\end{subequations}
Using a bit of algebra we can write the beam-splitter Hamiltonian in terms of Pauli operators as ${H}^{BS}_{ij}=X_{i}X_{j}+Y_{i}Y_{j}$. In \cite{Kempe2002}, Kempe and Whaley showed that this interaction acting between nearest and next-nearest neighbours on a 1D chain, along with Pauli $Z$ rotations, can perform universal quantum computation in an encoded subspace of the Fock space (next-to-nearest neighbour interactions are essential for the protocol to work). This result suggest that qubit linear optics is computationally more powerful than fermionic linear optics, even though the only differences are the signs of commutators of operators in different sites.

\subsection{The definition of fermionic anyons}\label{anyon}
The model we consider in this paper has been studied previously in the theory of interacting bosonic one-dimensional gases \cite{Amico1998One-dimensionalStatistics,Girardeau2006}, interacting fermionic one-dimensional gases \cite{Hao2012} and their simulations in several physical systems \cite{Keilmann2011,Greschner2015,Strater2016FloquetLattice,Cardarelli2016}. In this model the creation and annihilation operators ${a}^{\dagger}_{i}$ and ${a}_{i}$ satisfy the deformed anti-commutation relations
\begin{subequations}\label{eq13}
\begin{align}
&{a}_{i}{a}^{\dagger}_{j}+e^{-i\varphi\epsilon_{ij}}{a}^{\dagger}_{j}{a}_{i}=\delta_{i,j},\\
&{a}_{i}{a}_{j}+e^{i\varphi\epsilon_{ij}}{a}_{j}{a}_{i}=0,
\end{align}
where the symbol $\epsilon_{ij}$ is given by
\begin{equation}
\epsilon_{ij}=
\begin{cases}
\text{  }1&\text{, if }i<j\\
\text{  }0&\text{, if }i=j\\
-1&\text{, if }i>j
\end{cases}
,
\end{equation}
\end{subequations}
The dependence of the anti-commutation relations on $\epsilon_{ij}$ defines an order over the lattice, coming from the way the deformation is defined [see Eqs.\ (\ref{J-W})] . When $\varphi=0\text{ or }\pi$ this order is irrelevant and we re-obtain a fermionic system [see Eqs.\ (\ref{eq1})] and hard-core bosons [or qubits, see Eqs.\ (\ref{eq11d}) and (\ref{eq11e})], respectively. For all $0<\varphi<\pi$ we have a non-trivial anyonic model. The deformed anti-commutation relations come from the generalized Jordan-Wigner transformation below
\begin{subequations}\label{J-W}
\begin{align}
{a}^{\dagger}_{i}&={{JW}^{(\varphi)}_{i}}^{\dagger}{f}^{\dagger}_{i} ,\\
{a}_{i}&={JW}^{(\varphi)}_{i}{f}_{i},\\
{JW}^{(\varphi)}_{i}&=\exp{i\varphi\sum^{i-1}_{k=1}{f}^{\dagger}_{k}{f}_{k}}.
\end{align}
\end{subequations}
The Jordan-Wigner operator ${JW}^{(\varphi)}_{i}$ transmutes statistics in a way similar to the Chern-Simons field \cite{Lerda1993}.

Number operators for this system are as in the previous cases, and we represent them by ${N}^{A}_{i}={a}^{\dagger}_{i}{a}_{i}$ which have eigenvalues ${n}^{A}_{i}$ either $0$ or $1$. The Fock basis states for anyons are, therefore, 
\begin{equation}
\ket{{n}^{A}_{1},...,{n}^{A}_{m}}=({a}^{\dagger}_{1})^{{n}^{A}_{1}}...({a}^{\dagger}_{m})^{{n}^{A}_{m}}\ket{0_{A}}.
\end{equation}
Anyonic passive linear optical elements are the unitaries
\begin{subequations}
\begin{align}
{PS}_{i}(\theta)&=\exp{i\theta {H}^{PS}_{i}},\\
{BS}_{ij}(\theta)&=\exp{i\theta {H}^{BS}_{ij}},
\end{align}
with Hamiltonians given by
\begin{align}
{H}^{PS}_{i}&={N}^{A}_{i},\\
{H}^{BS}_{ij}&={a}^{\dagger}_{i}{a}_{j}+{a}^{\dagger}_{j}{a}_{i}.
\end{align}
\end{subequations}
The computational model is as before: the computational basis states are the Fock states, the unitaries are all possible combinations of phase-shifters and beam-splitters for arbitrary pairs of modes (this arbitrariness is essential for our result) and measurements are allowed in the Fock basis only.

\section{Linear optics of 1D fermionic anyons}\label{anyon1}
In this section we define the dynamics corresponding to anyonic linear optics. In section \ref{motion}, we solve the equations of motion for creation operators acted on by the linear-optical Hamiltonians exactly and explain the difference in relation to the fermionic case. Next, in section \ref{bohm} we show that when beam-splitters act on non-neighboring modes the presence of particles in between give rise to an Aharonov-Bohm effect, and we argue that this is responsible for the differences in the action of beam-splitters with respect to free fermions.

\subsection{Anyonic optical elements and equations of motion}\label{motion}

In the anyonic case, the evolution of creation operators is more involved than for the bosonic and fermionic cases, and it is what we discuss now. For phase-shifters we have that
\begin{equation}
{PS}_{i}(\theta){a}^{\dagger}_{j}{PS}_{i}(-\theta)=e^{i\theta\delta_{ij}}{a}^{\dagger}_{j},
\end{equation}
just like before. The action of beam-splitters, on the other hand, requires more attention. We will solve the dynamical problem defined by the ${H}^{BS}_{ij}$ hamiltonian. The Heisenberg equations of motion are
\begin{subequations}
\begin{align}
i\frac{d{a}^{\dagger}_{i}}{d\theta}=[{H}^{BS}_{ij},{a}^{\dagger}_{i}]&,\\
i\frac{d{a}^{\dagger}_{j}}{d\theta}=[{H}^{BS}_{ij},{a}^{\dagger}_{j}]&,
\end{align}
\end{subequations}
where the commutators are computed using algebra of anyonic operators [Eqs.\ (\ref{eq13})]:
\begin{subequations}
\begin{align}
\left[{H}^{BS}_{ij},{a}^{\dagger}_{i}\right]&={a}^{\dagger}_{j}\{1-(1-e^{i\varphi}){N}^{A}_{i}\},\\ 
\left[{H}^{BS}_{ij},{a}^{\dagger}_{j}\right]&={a}^{\dagger}_{i}\{1-(1-e^{-i\varphi}){N}^{A}_{j}\}.
\end{align}
\end{subequations}

Let us start by rewriting the equation for mode $i$ in a more suggestive form:
\begin{equation}
i\frac{d{a}^{\dagger}_{i}}{d\theta}={a}^{\dagger}_{j}{W^{(\varphi)}_{i}},
\end{equation}
where we have introduced a short-hand notation for the non-linear term ${W^{(\varphi)}_{i}}\equiv 1-(1-e^{i\varphi}){N}^{A}_{i}$. By computing the commutator $[{H}^{BS}_{ij},{a}^{\dagger}_{j}{W^{(\varphi)}_{i}}]$ we find that the equation of motion for this operator is given by
\begin{equation}
i\frac{d({a}^{\dagger}_{j}{W^{(\varphi)}_{i}})}{d\theta}={a}^{\dagger}_{i},
\end{equation}
therefore this is a coupled linear system of equations for the operators ${a}^{\dagger}_{i}$ and ${a}^{\dagger}_{j}{W^{(\varphi)}_{i}}$, which is exactly solvable. The equation of motion for mode $j$ has a similar property giving the system
\begin{subequations}
\begin{align}
i\frac{d{a}^{\dagger}_{j}}{d\theta}={a}^{\dagger}_{i}{{W^{\dagger}}^{(\varphi)}_{j}}&,\\
i\frac{d({a}^{\dagger}_{i}{W^{\dagger}}^{(\varphi)}_{j})}{d\theta}={a}^{\dagger}_{j}.&
\end{align}
\end{subequations}
By linearity, the solutions of the equations for ${a}^{\dagger}_{i}$ and ${a}^{\dagger}_{j}$ must be:
\begin{subequations}
\begin{align}
&{BS}_{ij}(\theta){a}^{\dagger}_{i}{BS}_{ij}(-\theta)=\cos{\theta}{a}^{\dagger}_{i}+i\sin{\theta}{a}^{\dagger}_{j}{W^{(\varphi)}_{i}},\label{23a}\\
&{BS}_{ij}(\theta){a}^{\dagger}_{j}{BS}_{ij}(-\theta)=\cos{\theta}{a}^{\dagger}_{j}+i\sin{\theta}{a}^{\dagger}_{i}{W^{\dagger}}^{(\varphi)}_{j}.\label{23b}
\end{align}
\end{subequations}
Notice that these solutions are very similar to the corresponding dynamics of fermions and bosons [Eqs.(\ref{eq5}) and (\ref{eq9b})].

Eqs.(\ref{23a}) and (\ref{23b}) represent the exact dynamics for arbitrary pairs of modes for these systems. Previous treatments have focused on many-body anyonic Hamiltonians (for either bosonic and fermionic anyon varieties) using methods such as the Bethe Ansatz, as in  \cite{Amico1998One-dimensionalStatistics}.

\subsection{Aharonov-Bohm effect in anyonic beam-splitters}\label{bohm}

To complete our description of a general beam-splitter we need to discuss what happens with the modes that are between the ones appearing in the Hamiltonian. In the fermionic and bosonic cases, these modes commute with $H^{BS}$, but in the anyonic case creation operators ${a}^{\dagger}_{k}$ with $i<k<j$ do not commute with $H^{BS}_{ij}$. In fact, they satisfy the relation
\begin{subequations}
\begin{align}
&\exp{i\theta({a}^{\dagger}_{i}{a}_{j}+{a}^{\dagger}_{j}{a}_{i})}{a}^{\dagger}_{k}=\\
&={a}^{\dagger}_{k}\exp{i\theta(e^{i2\varphi}{a}^{\dagger}_{i}{a}_{j}+e^{-i2\varphi}{a}^{\dagger}_{j}{a}_{i})},
\end{align}
\end{subequations}
which we can treat as an effective beam-splitter that acts on states by introducing a phase correction dependent on the number of modes occupied between $i$ and $j$. Or putting it in a more compact notation
\begin{subequations}\label{eq26}
\begin{align}
{BS}_{ij}(\theta){a}^{\dagger}_{k}={a}^{\dagger}_{k}{BS}_{ij}^{(2\varphi)}(\theta),\label{eq26a}  
\end{align}
with the new effective beam-splitter unitary defined by
\begin{align}
{BS}_{ij}^{(\alpha)}(\theta)=\exp{i\theta(e^{i\alpha}{a}^{\dagger}_{i}{a}_{j}+e^{-i\alpha}{a}^{\dagger}_{j}{a}_{i})},
\end{align}
and the solution to the equations of motion for modes $i$ and $j$ is:
\begin{align}
{a^{(\alpha)}}^{\dagger}_{i}(\theta)&=\cos{\theta}{a}^{\dagger}_{i}+ie^{i\alpha}\sin{\theta}{a}^{\dagger}_{j}{W^{(\varphi)}_{i}},\label{eq26c}\\
{a^{(\alpha)}}^{\dagger}_{j}(\theta)&=\cos{\theta}{a}^{\dagger}_{j}-ie^{-i\alpha}\sin{\theta}{a}^{\dagger}_{i}{W^{\dagger}}^{(\varphi)}_{j}.\label{eq26d}
\end{align}
\end{subequations}

\begin{exmp*}\label{exmp}
To illustrate this phase correction, consider a system with $3$ modes, and choose $i=1$ and $j=3$. Let us act with a balanced beam-splitter $(\theta=\frac{\pi}{4})$ on state $\ket{0,1,1}={a}^{\dagger}_{2}{a}^{\dagger}_{3}\ket{0_{A}}$. The first step of the calculation is to use Eq. (\ref{eq26a}):
\begin{equation}
{BS}_{13}\left(\frac{\pi}{4}\right){a}^{\dagger}_{2}{a}^{\dagger}_{3}\ket{0_{A}}={a}^{\dagger}_{2}\left[{BS}^{(2\varphi)}_{13}\left(\frac{\pi}{4}\right)\right]{a}^{\dagger}_{3}\ket{0_{A}}.
\end{equation}
Now we need to analyze the dynamics of ${a}^{\dagger}_{3}\ket{0_{A}}$ under the action of ${BS}^{(2\varphi)}_{13}\left(\frac{\pi}{4}\right)$. Using Eqs. (\ref{eq26c}) and (\ref{eq26d}), and the fact that $W^{(\varphi)}_{i}\ket{0_{A}}=\ket{0_{A}}$ for every mode $i$ and angle $\varphi$ we obtain:
\begin{equation}
{a}^{\dagger}_{2}\left[{BS}^{(2\varphi)}_{13}\left(\frac{\pi}{4}\right)\right]{a}^{\dagger}_{3}\ket{0_{A}}=\frac{1}{\sqrt{2}}{a}^{\dagger}_{2}(ie^{i2\varphi}{a}^{\dagger}_{1}+{a}^{\dagger}_{3})\ket{0_{A}}.
\end{equation}
We now use the commutation relation ${a}^{\dagger}_{2}{a}^{\dagger}_{1}=-e^{-i\varphi}{a}^{\dagger}_{1}{a}^{\dagger}_{2}$ to write the final equation in normally ordered form
\begin{equation}
\frac{1}{\sqrt{2}}{a}^{\dagger}_{2}(ie^{i2\varphi}{a}^{\dagger}_{1}+{a}^{\dagger}_{3})\ket{0_{A}}=\frac{1}{\sqrt{2}}(-ie^{i\varphi}{a}^{\dagger}_{1}{a}^{\dagger}_{2}+{a}^{\dagger}_{2}{a}^{\dagger}_{3})\ket{0_{A}}.
\end{equation}
If we compare this result with the same Hamiltonian applied on state ${a}^{\dagger}_{1}{a}^{\dagger}_{2}\ket{0_{A}}$, we see that the effect of the phase-corrected Hamiltonian is to guarantee unitarity of the time evolution operator. To see this use the solution of the equations of motion to obtain
\begin{equation}
{BS}_{13}\left(\frac{\pi}{4}\right){a}^{\dagger}_{1}{a}^{\dagger}_{2}\ket{0_{A}}=\frac{1}{\sqrt{2}}({a}^{\dagger}_{1}+i{a}^{\dagger}_{3}){a}^{\dagger}_{2}\ket{0_{A}}.
\end{equation}
Finally, using ${a}^{\dagger}_{3}{a}^{\dagger}_{2}=-e^{-i\varphi}{a}^{\dagger}_{2}{a}^{\dagger}_{3}$ we get
\begin{equation}
\frac{1}{\sqrt{2}}({a}^{\dagger}_{1}+i{a}^{\dagger}_{3}){a}^{\dagger}_{2}\ket{0_{A}}=\frac{1}{\sqrt{2}}({a}^{\dagger}_{1}{a}^{\dagger}_{2}-ie^{-i\varphi}{a}^{\dagger}_{2}{a}^{\dagger}_{3})\ket{0_{A}}.
\end{equation}
Therefore the total effect on the Fock basis is given by
\begin{subequations}
\begin{align}
{BS}_{13}\left(\frac{\pi}{4}\right)\ket{1,1,0}&=\frac{1}{\sqrt{2}}(\ket{1,1,0}-ie^{-i\varphi}\ket{0,1,1}),\\
{BS}_{13}\left(\frac{\pi}{4}\right)\ket{0,1,1}&=\frac{1}{\sqrt{2}}(-ie^{i\varphi}\ket{1,1,0}+\ket{0,1,1}),
\end{align}
\end{subequations}
which is manifestly unitary. 
\end{exmp*}
If the anyon in mode $1$ tunnels to mode $3$ when another anyon occupies mode $2$, a relative phase factor between the modes appears. It is easy to see that this phase would be absent if mode $2$ was empty. This dynamics can be understood as an one-dimensional analogue of the Aharonov-Bohm effect \cite{Aharonov1959SignificanceTheory}, where the magnetic flux carried by the particle is given by $\pi-\varphi$ and the $\pi$ factor account for the $-1$ fermionic phase appearing when $\varphi=0$.

This phenomenon, due to particle statistics, was seen previously in  \cite{HaugTheCondensates}, where it affects the phase-difference of a driven Bose-Einstein condensate (of bosonic anyons) in ring-shaped optical lattices threaded by a magnetic flux.  These phase factors, as we will see in the next section, are crucial in determining the computational power of the different models.

\section{The computational power of anyons}\label{power}

In \cite{Kempe2002}, it was shown how to construct an encoded, entangling two-qubit gate using only nearest and next-to-nearest neighbour Hamiltonians ($XY$ interaction) between physical qubits, to achieve universal quantum computation. In this section we will generalize this construction to show that quadratic dynamics of fermionic anyons is also capable of universal quantum computation. We begin by defining a qubit encoding that is preserved by dynamics of either free fermions, hard-core bosons, or free fermionic anyons. We will then describe linear-optical circuits on fermionic anyons and its action on encoded states. We prove that the two-qubit logical gate implemented by this circuit is deterministic and entangling for any value of the statistical parameter $\varphi\neq 0$. This, together with single-qubit unitaries, shows that linear optics on fermionic anyons is universal for quantum computation.

\subsection{Encoding}
We use $2n$ modes to encode $n$ qubits such that each logical qubit corresponds to a pair of neighboring modes as in the equations
\begin{subequations}
\begin{align}
\ket{0_{L}}&=\ket{1,0}\\
\ket{1_{L}}&=\ket{0,1}
\end{align}
\end{subequations}
So, for example, a two-qubit system needs four modes and the logical states are given by
\begin{subequations}
\begin{align}
\ket{00}_{L}&=\ket{1,0,1,0},\\
\ket{01}_{L}&=\ket{1,0,0,1},\\
\ket{10}_{L}&=\ket{0,1,1,0},\\
\ket{11}_{L}&=\ket{0,1,0,1},
\end{align}
\end{subequations}
where the right-hand-side of these equations are Fock states. This encoding is independent of the parameter $\varphi$, since all of these particles obey Pauli Exclusion Principle, which allows the direct comparison of logical gates between models.

\subsection{Encoded one- and two-qubit gates}
\begin{figure}[h!]
\centering
\includegraphics[scale=0.9]{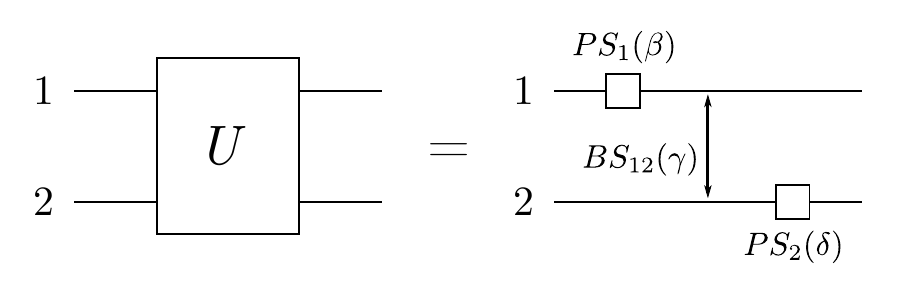}
\caption{Single-qubit unitary decomposed in optical elements: A single-qubit unitary needs four parameters ($\alpha,\beta,\gamma,\delta$). The first one is a global phase and the others are realized by the optical elements in the figure.}
\label{fig:unitary}
\end{figure}

With this encoding it is possible to do any logical one-qubit gate using only phase-shifters and beam-splitters on the two corresponding modes. To prove this, consider a single qubit encoded in modes 1 and 2, and notice that a phase-shifter in mode 2 acts in the logical basis states as
\begin{subequations}
\begin{align}
&{PS}_{2}(\theta)\ket{1,0}=\ket{1,0},\\
&{PS}_{2}(\theta)\ket{0,1}=e^{i\theta}\ket{0,1},
\end{align}
\end{subequations}
which is equivalent to a logical $Z$ rotation in the Bloch sphere by $\theta$ degrees. Note also that a beam-splitter between modes 1 and 2 acts in the logical basis states as
\begin{subequations}
\begin{align}
&{BS}_{12}(\theta)\ket{1,0}=\cos{\theta}\ket{1,0}+i\sin{\theta}\ket{0,1},\\
&{BS}_{12}(\theta)\ket{0,1}=i\sin{\theta}\ket{1,0}+\cos{\theta}\ket{0,1},
\end{align}
\end{subequations}
which is equivalent to a logical $X$ rotation in the Bloch sphere, by an angle $\theta$. With arbitrary rotations around two axes in the Bloch sphere, we can perform an arbitrary single-qubit gate, as in Fig.\ref{fig:unitary}.

\begin{figure}[htbp]
\centering
\includegraphics[scale=0.9]{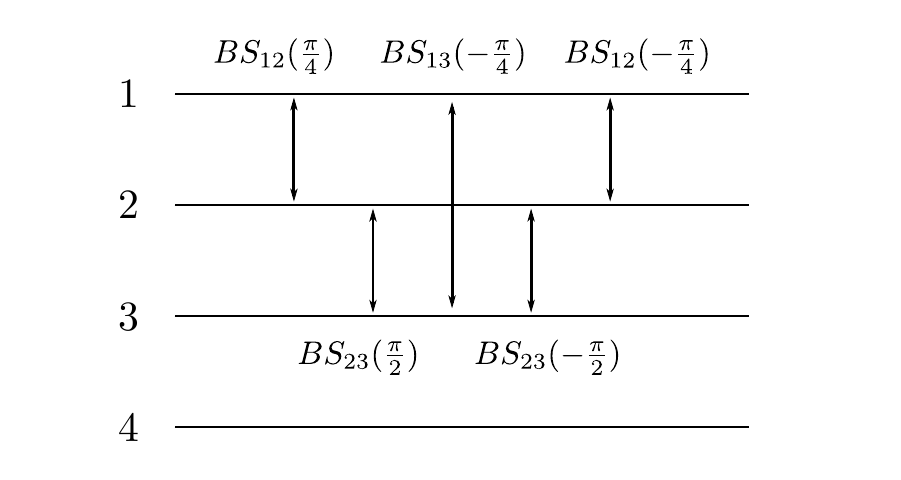}
\caption{Two-qubit gate: sequence of beam-splitters that generate our entangling gate}
\label{fig:circuit}
\end{figure}

To implement an encoded two-qubit gate we generalize the XY-interaction protocol found in \cite{Kempe2002} (see Fig.\ \ref{fig:circuit}), adapting the construction to our anyonic model.

When the particles considered are hard-core bosons (or spins), we recover the original result of \cite{Kempe2002} and the circuit executes the logical gate $\sqrt{-ZZ}$ which is a maximally entangling gate. %sufficient to complement single-qubit gates and achieve universal quantum computation. 

In Appendix A we calculate the effect of this circuit on fermionic anyons characterized by the deformed fermionic algebra of Eqs.(\ref{eq13}), for any value of the deformation parameter $\varphi$. Its action on the encoded qubits is the gate
\begin{equation}
{C}(\varphi)=R_{\hat{z}}(\frac{\pi}{2})\otimes\ketbra{0}+R_{{\hat{n}}}(\frac{\pi}{2})\otimes\ketbra{1}, \label{eq:cphi}
\end{equation}
where $R_{\hat{z}}(\frac{\pi}{2})$ is a rotation of $\frac{\pi}{2}$ around the $Z$ axis in the Bloch sphere, and $R_{{\hat{n}}}(\frac{\pi}{2})$ is a $\frac{\pi}{2}$ rotation around the axis ${\hat{n}}=(-\sin{\varphi},0,\cos{\varphi})$ in the Bloch sphere. So this is a controlled rotation gate whose action depends continuously on the parameter $\varphi$ characterizing our anyonic model. We have, for the special cases of fermions and hard-core bosons,
\begin{equation}
C(\varphi)=
\begin{cases}
Z\otimes\mathbf{1}&\text{, if }\varphi=0\\
\sqrt{-ZZ}&\text{, if }\varphi=\pi
\end{cases}
,
\end{equation}
which in the fermionic case is a local gate, and for hard-core bosons is as described above. Therefore, this gate interpolates between the corresponding ones for the other two models.

\subsection{The entangling power of ${C}(\varphi)$}

We claim that gate ${C}(\varphi)$ [Eq.(\ref{eq:cphi})], together with arbitrary single-qubit gates, form a set that is universal for quantum computing whenever $\varphi\neq 0$. To prove this it is sufficient to show that ${C}(\varphi)$ has a non-zero entangling power (as defined by \cite{Zanardi2000}).

The entangling power $e_{p}(U)$ of a unitary gate $U$ is defined as the average entanglement of formation generated by the action of $U$ on product states $\ket{\psi_{1}}\otimes\ket{\psi_{2}}$
\begin{equation}
e_{p}(U)=\overline{E(U\ket{\psi_{1}}\otimes\ket{\psi_{2}})}^{(\psi_{1},\psi_{2})},
\end{equation}
where the bar denotes average with respect to some probability distribution $p(\psi_{1},\psi_{2})$. It can be shown that, if the average is taken over the uniform distribution, the entangling power is both local invariant and \textsc{SWAP} invariant (that is, it remains the same if $U$ is conjugated by \textsc{SWAP} or by single-qubit gates).
In fact, this invariant can be easily calculated in terms of simpler invariants, which was done in \cite{Balakrishnan2010}. Two-qubit gates have two local invariant quantities, given by
\begin{equation}
G_{1}=\frac{\Tr^{2}{U^{T}_{B}U_{B}}}{16\det(U)},
\end{equation}
and
\begin{equation}
G_{2}=\frac{\Tr^{2}{U^{T}_{B}U_{B}}-\Tr{(U^{T}_{B}U_{B})^{2}}}{4\det(U)},
\end{equation}
where $U_{B}$ is the matrix representation of the gate $U$ written in the \textit{Bell basis}. With these invariants, the (normalized) entangling power $e_{p}(U)$ of a two-qubit gate $U$ over the uniform distribution is just given by \cite{Balakrishnan2010}
\begin{equation}
e_{p}(U)=1-\abs{G_{1}}. 
\end{equation}
With this in hands, the entangling power of ${C}(\varphi)$ [Eq. (\ref{eq:cphi})] is:
\begin{equation}
e_{p}({C}(\varphi))=1-\cos^{4}{\frac{\varphi}{2}}.
\end{equation}
This shows that any $\varphi \neq 0$ results in a fermionic anyon model that allows for universal quantum computation. In fact, since this gate generates entanglement, it can be used to construct encoded \textsc{CNOT} gates using the argument given in \cite{Bremner2002}, with the number of required ${C}(\varphi)$ gates depending on the value of $e_{p}({C}(\varphi))$.
\\
\section{Conclusion} \label{conclusion}
We generalized the model of passive Fermionic Linear Optics by studying anyonic systems defined by deformations of fermionic anti-commutation relations. We have taken quadratic, number-preserving Hamiltonians as the analogue of fermionic linear-optical dynamics and solved the generated dynamics exactly. We showed how the difference to fermionic dynamics is due to a one-dimensional analogue of the Aharonov-Bohm effect, which is intimately related to the computational power of the model we consider, and also gives a simple interpretation to a previously known effect \cite{HaugTheCondensates}.

We adapted a scheme for quantum computing with nearest- and next-nearest-neighbour spin-$\frac{1}{2}$ interactions to systems of identical particles, showing that the analogous interactions in our model allow for universal quantum computation when the particles have fractional exchange statistics. This happens for any value of the statistical deformation parameter $\varphi \neq 0$, that is, as long as our anyons differ from usual fermions. Given that free fermions can be simulated efficiently, this means the transition in computational power is abrupt, as is common to extensions of FLO \cite{Brod2011,Oszmaniec2017UniversalOperations}, going from classically simulable to quantum universal for any $\varphi \neq 0$.

This raises the question of whether such interactions arising from statistics alone can lead to a computational advantage in other settings. For example, it would be interesting to study deformed bosonic commutation relations. This is likely to be a harder problem than the fermionic one, due to the absence of an exclusion principle, which simplifies the dynamical equations. Deformed bosonic relations would result in a generalization of the BosonSampling model. BosonSampling is known to be hard to simulate classically (given some complexity-theoretic assumptions), but is not known to be universal for quantum computation. We leave the exact dynamics of two-body hopping Hamiltonians for anyonic bosons as an interesting open problem.

The fact that the fermionic anyons we consider allow for universal quantum computation can also be interpreted as a statement about the complexity of the model's dynamics. That is, since we proved the universality of the model, this provides strong evidence that it has no efficient classical simulation scheme. In addition, no quantum computational device that is not universal will be able to efficiently simulate fermionic anyons. Thus, our results shows how studying the computational capacity of general anyonic models may yield insights on their complexity.

\section{Acknowledgements}
This work was supported by project Instituto Nacional de Ciência e Tecnologia de Informação Quântica (INCT-IQ/CNPq).
\bibliographystyle{apsrev4-1}
\bibliography{references.bib}
\appendix
\section{Explicit expression of ${C}(\varphi)$}

In this Appendix we calculate the two-qubit unitary implemented by the anyonic linear-optical circuit of Fig. \ref{fig:circuit}. For this, we need to calculate the matrix elements of the various beam-splitters in the two-particle basis, and see if our protocol preserves the encoding of qubits. Given the four modes of Fig. \ref{fig:circuit} the two-particle basis is given by the states
\begin{subequations}
\begin{align}
&\ket{1,1,0,0},\ket{1,0,1,0},\ket{1,0,0,1},\\
&\ket{0,1,1,0},\ket{0,1,0,1},\ket{0,0,1,1},
\end{align}
\end{subequations}
There are three kinds of beam-splitters in the circuit, ${BS}_{12},{BS}_{23}$ and ${BS}_{13}$ where the first two appear more then once, with two different angles (defining the beam-splitting ratios). First let us calculate the matrix representation of the general beam-splitter ${BS}_{12}(\theta)$ in our two-particle basis. The idea is to use the representation in terms of creation operators and use the solutions to the equations of motion to find the matrix elements. For brevity, we will do the calculation for the more involved basis states, as the others will be similar in form, and easier to calculate.

The first case will be ${a}^{\dagger}_{1}{a}^{\dagger}_{2}\ket{0_{A}}$,
\begin{equation}
\begin{aligned}
&{BS}_{12}(\theta){a}^{\dagger}_{1}{a}^{\dagger}_{2}\ket{0_{A}}=\\
&=[{BS}_{12}(\theta){a}^{\dagger}_{1}{BS}_{12}(-\theta)]{BS}_{12}(\theta){a}^{\dagger}_{2}\ket{0_{A}},
\end{aligned}
\end{equation}
where ${BS}_{12}(\theta)\ket{0_{A}}=\ket{0_{A}}$. We can now use the solutions to the equations of motion to obtain
\begin{equation}
\begin{aligned}
&[{BS}_{12}(\theta){a}^{\dagger}_{1}{BS}_{12}(-\theta)]{BS}_{12}(\theta){a}^{\dagger}_{2}\ket{0_{A}}=\\
&=(\cos{\theta}{a}^{\dagger}_{1}+i\sin{\theta}{a}^{\dagger}_{2}{W^{(\varphi)}_{1}})\times\\
&\times(\cos{\theta}{a}^{\dagger}_{2}+i\sin{\theta}{a}^{\dagger}_{1}{W^{\dagger}}^{(\varphi)}_{2})\ket{0_{A}},
\end{aligned}
\end{equation}
Using that $[{W}^{(\varphi)}_{1},{a}^{\dagger}_{2}]=0$ we can rewrite this as:
\begin{equation}
\cos^{2}{\theta}{a}^{\dagger}_{1}{a}^{\dagger}_{2}\ket{0_{A}}-\sin^{2}{\theta}{a}^{\dagger}_{2}{W}^{(\varphi)}_{1}{a}^{\dagger}_{1}\ket{0_{A}},
\end{equation}
in the next step, we use the identity ${N}^{A}_{1}{a}^{\dagger}_{1}={a}^{\dagger}_{1}$ to obtain
\begin{equation}
\begin{aligned}
&\cos^{2}{\theta}{a}^{\dagger}_{1}{a}^{\dagger}_{2}\ket{0_{A}}-\sin^{2}{\theta}{a}^{\dagger}_{2}[1-(1-e^{i\varphi}){N}^{A}_{1}]{a}^{\dagger}_{1}\ket{0_{A}}=\\
&=\cos^{2}{\theta}{a}^{\dagger}_{1}{a}^{\dagger}_{2}\ket{0_{A}}-e^{i\varphi}\sin^{2}{\theta}{a}^{\dagger}_{2}{a}^{\dagger}_{1}\ket{0_{A}},
\end{aligned}
\end{equation}
Finally, using the commutation relation ${a}^{\dagger}_{1}{a}^{\dagger}_{2}=-e^{i\varphi}{a}^{\dagger}_{2}{a}^{\dagger}_{1}$ we conclude that
\begin{equation}
\begin{aligned}
&{BS}_{12}(\theta){a}^{\dagger}_{1}{a}^{\dagger}_{2}\ket{0_{A}}=\\
&=\cos^{2}{\theta}{a}^{\dagger}_{1}{a}^{\dagger}_{2}\ket{0_{A}}-e^{i\varphi}\sin^{2}{\theta}{a}^{\dagger}_{2}{a}^{\dagger}_{1}\ket{0_{A}}=\\
&=(\cos^{2}{\theta}+\sin^{2}{\theta}){a}^{\dagger}_{1}{a}^{\dagger}_{2}\ket{0_{A}}={a}^{\dagger}_{1}{a}^{\dagger}_{2}\ket{0_{A}},
\end{aligned}
\end{equation}
So the matrix element $\bra{1,1,0,0}{BS}_{12}(\theta)\ket{1,1,0,0}$ is $1$. Similarly $\bra{0,0,1,1}{BS}_{12}(\theta)\ket{0,0,1,1}$ is also $1$, because the beam-splitter has no action on these modes. Now we illustrate one more case, and then give the expression for ${BS}_{12}(\theta)$. Consider the state ${a}^{\dagger}_{1}{a}^{\dagger}_{3}\ket{0_{A}}$, we can proceed in pretty much the same way we did before, and obtain
\begin{equation}
\begin{aligned}
&{BS}_{12}(\theta){a}^{\dagger}_{1}{a}^{\dagger}_{3}\ket{0_{A}}=\\
&=[{BS}_{12}(\theta){a}^{\dagger}_{1}{BS}_{12}(-\theta)]{BS}_{12}(\theta){a}^{\dagger}_{3}\ket{0_{A}}=\\
&=[{BS}_{12}(\theta){a}^{\dagger}_{1}{BS}_{12}(-\theta)]{a}^{\dagger}_{3}\ket{0_{A}}=\\
&=(\cos{\theta}{a}^{\dagger}_{1}+i\sin{\theta}{a}^{\dagger}_{2}{W}^{(\varphi)}_{1}){a}^{\dagger}_{3}\ket{0_{A}}=\\
&=\cos{\theta}{a}^{\dagger}_{1}{a}^{\dagger}_{3}\ket{0_{A}}+i\sin{\theta}{a}^{\dagger}_{2}{a}^{\dagger}_{3}\ket{0_{A}},
\end{aligned}
\end{equation}
which tells us that $\bra{1,0,1,0}{BS}_{12}(\theta)\ket{1,0,1,0}=\cos{\theta}$ and $\bra{0,1,1,0}{BS}_{12}(\theta)\ket{1,0,1,0}=i\sin{\theta}$. Doing the calculation of the other matrix elements we obtain
\begin{equation}
[{BS}_{12}(\varphi)]=
\begin{bmatrix}
1 & 0 & 0 & 0 & 0 & 0\\
0 & \cos{\theta} & 0 & i\sin{\theta} & 0 & 0\\
0 & 0 & \cos{\theta} & 0 & i\sin{\theta} & 0\\
0 & i\sin{\theta} & 0 & \cos{\theta} & 0 & 0\\
0 & 0 & i\sin{\theta} & 0 & \cos{\theta} & 0\\
0 & 0 & 0 & 0 & 0 & 1\\
\end{bmatrix}
.
\end{equation}
The matrix for ${BS}_{23}(\theta)$ is very similar, since most of the calculations of matrix are elements repeated with different indices. The matrix is
\begin{equation}
[{BS}_{23}(\theta)]=
\begin{bmatrix}
\cos{\theta} & i\sin{\theta} & 0 & 0 & 0 & 0\\
i\sin{\theta} & \cos{\theta} & 0 & 0 & 0 & 0\\
0 & 0 & 1 & 0 & 0 & 0\\
0 & 0 & 0 & 1 & 0 & 0\\
0 & 0 & 0 & 0 & \cos{\theta} & i\sin{\theta}\\
0 & 0 & 0 & 0 & i\sin{\theta} & \cos{\theta}\\
\end{bmatrix}
.
\end{equation}

The remaining matrix is the one which marks a departure between fermions and anyons, as the 1D analogue of  the Aharonov-Bohm phase appears explicitly. To calculate it we must use the result of Example \ref{exmp} for the matrix elements $\bra{1100}{BS}_{13}(\theta)\ket{1100}$, $\bra{0110}{BS}_{13}(\theta)\ket{1100}$, $\bra{1100}{BS}_{13}(\theta)\ket{0110}$, and $\bra{0110}{BS}_{13}(\theta)\ket{0110}$. The other matrix elements are trivial. The result of this calculation is
\begin{equation}
[{BS}_{13}(\theta)]=
\begin{bmatrix}
\cos{\theta} & 0 & 0 & i\sin{\theta} & 0 & 0\\
0 & 1 & 0 & 0 & 0 & 0\\
0 & 0 & \cos{\theta} & 0 & 0 & -ie^{-i\varphi}\sin{\theta}\\
i\sin{\theta} & 0 & 0 & \cos{\theta} & 0 & 0\\
0 & 0 & 0 & 0 & 1 & 0\\
0 & 0 & -ie^{i\varphi}\sin{\theta} & 0 & 0 & \cos{\theta}\\
\end{bmatrix}
,
\end{equation}
Now, combining all of these results to evaluate the matrix products indicated by the circuit (Fig.\ref{fig:circuit}) we obtain the matrix
\begin{equation}
[{C}(\varphi)]=
\begin{bmatrix}
1 & 0 & 0 & 0 & 0 & 0\\
0 & e^{-i\frac{\pi}{4}} & 0 & 0 & 0 & 0\\
0 & 0 & \frac{1-i\cos{\varphi}}{\sqrt{2}} & 0 & \frac{i\sin{\varphi}}{\sqrt{2}} & 0\\
0 & 0 & 0 & e^{i\frac{\pi}{4}} & 0 & 0\\
0 & 0 & \frac{i\sin{\varphi}}{\sqrt{2}} & 0 & \frac{1+i\cos{\varphi}}{\sqrt{2}} & 0\\
0 & 0 & 0 & 0 & 0 & 1\\
\end{bmatrix}
,
\end{equation}
To stay in the encoding defined in section \ref{power}, the image of the encoded states under the unitary must stay in the encoded subspace, and to guarantee this we only need to show that $\ket{1,1,0,0}$ and $\ket{0,0,1,1}$ are eigenstates of ${C}(\varphi)$, which is easily seen in the matrix above. In fact ${C}(\varphi)$ in the encoded basis is 

\begin{equation}
\begin{bmatrix}
e^{-i\frac{\pi}{4}} & 0 & 0 & 0\\
0 & \frac{1-i\cos{\varphi}}{\sqrt{2}} & 0 & \frac{i\sin{\varphi}}{\sqrt{2}}\\
0 & 0 & e^{i\frac{\pi}{4}} & 0\\
0 & \frac{i\sin{\varphi}}{\sqrt{2}} & 0 & \frac{1+i\cos{\varphi}}{\sqrt{2}}\\
\end{bmatrix}
,
\end{equation}
showing that ${C}(\varphi)=R_{\hat{z}}(\frac{\pi}{2})\otimes\ketbra{0}+R_{\hat{n}}(\frac{\pi}{2})\otimes\ketbra{1}$, with $\hat{n}=(-\sin{\varphi},0,\cos{\varphi})$ as claimed.
\end{document}